\documentclass[]{article}
\usepackage{amsmath,amssymb}
\usepackage{lmodern}
\usepackage{xcolor}
\usepackage[margin=1in]{geometry}
\usepackage{color}
\usepackage{fancyvrb}
\usepackage{graphicx}
\usepackage{hyperref}
\usepackage{microtype}

\title{R Package moodlequizR: Fully Randomized Moodle Tests.}
\author{by Wolfgang Rolke}

\begin{document}

\maketitle

\abstract{%
This article describes the \textbf{R} \cite{r2021} package \emph{moodlequizR}, which allows the user to easily create fully randomized quizzes and exams for Moodle, or indeed any online assessment platform that uses XML files for importing questions. In such a quiz the students are presented with the essentially same problem, but with various parts sufficiently different to make cheating very difficult. For example, the problem might require the students to find the sample mean but each student is presented with a different data set. Moodle does include some facilities for randomization, but these are rudimentary and wholly insufficient for a course in Statistics. The package is available on \href{https://cran.r-project.org/}{CRAN}.
}

\hypertarget{introduction}{%
\section{Introduction}\label{introduction}}

There has long been a need to move in-class assessments such as quizzes, home works and exams to some online platform. One reason for doing so is to lighten the load of the instructor by eliminating or at least lowering the time spent on grading papers. This of course becomes absolutely necessary for online courses, which have become much more wide-spread during and after the Covid pandemic. 

There has also been a steady increase in University courses that are fully designed to be held online, with the obvious need for online assessment. Finally these systems are also in use outside of academia, in industry as well as in government. For a detailed study on online teaching and learning see \cite{martin}.

There are at least 20 online teaching and assessment platforms in wider use, such as LearnWorlds, LinkedIn Learning, Udemy, Coursera, Canvas, OpenOlat, Blackboard and many more. One of the oldest and still (in the opinion of this author) best is \href{https://moodle.org/}{Moodle}. While \emph{moodlequizR} was written with Moodle in mind, it can in principle also be used with any assessment platform that uses the XML format to import questions, for example OpenOlat. 

\begin{sloppypar}
One concern with such online assessments is the ease with which students could cheat. \cite{amigud} gives a detailed study of why, how and when students are cheating in exams.  \cite{garg} provide a comprehensive review of the issues involved in assessment security. \cite{golden} discusses specifically the issue of security of test bank questions in online courses.  
\end{sloppypar}

One way to limit this danger is by presenting each student with a different (aka randomized) test. In fact Moodle does allow for some randomization. However, this is quite limited, for example different numbers can be chosen randomly but only from a uniform distribution. Certainly for a Statistics course this is insufficient. \emph{moodlequizR} allows the user to create randomized Moodle tests with a very high degree of randomization, far beyond simply choosing random numbers or shuffling the order of the questions. It also allows for the use of graphs in tests and it includes routines that make it easy for even beginning undergraduate students to transfer data from a Moodle test to \textbf{R}, so they can then use \textbf{R} to find the answers to the test questions.

Because Moodle tests are also graded automatically their use eliminates the need to grade papers by hand, and so eliminates a chore few instructors cherish. At the end of the semester the instructor can simply download the scores, for example as an Excel file, and then quickly assign a final grade to each student. There is even an \textbf{R} package called \emph{moodleR} \cite{moodleR} to help with the analysis of Moodle test scores.

The use of the package requires some knowledge of \textbf{R}. However, it comes with a shiny web application that allows even those with only a limited knowledge of \textbf{R} to create simple quizzes and exams.

While the package was written with courses in Statistics in mind, it can also be used to create Moodle tests for other fields. At the end of this article we will describe a test for a calculus course. It should be useful for any exams where the answers to questions generally are either numeric, multiple choice or very short texts. 

\hypertarget{other-packages}{%
\section{Similar \textbf{R} Packages}\label{other-packages}}

One existing package with a similar aim is \emph{exams} \cite{exams}. Like \emph{moodlequizR} it can be used to create randomized tests with \textbf{R}, which then can be imported into Moodle or similar learning systems. However, there are a number of differences. The shiny app packaged with \emph{moodlequizR} allows a new user to create a first test with almost no effort. \emph{moodlequizR} includes 15 template questions, covering a large number of the types of questions typical in a Statistics course. These can then be used as a starting point and changed to create other questions. On the other end of the learning curve by directly changing the \textbf{R} scripts a user has essentially unlimited ways to randomize the questions, much more than is possible with \emph{exams}.  

Another important difference is that with \emph{moodlequizR} randomized data can be embedded directly into the questions, something not possible with \emph{exams}. Students can then easily copy-paste the data into \textbf{R} with the included routine \emph{paste.data} and so use \textbf{R} to answer the questions.

The \emph{moostr} (\href{https://github.com/iris-yi-jiang/moostr}{https://github.com/iris-yi-jiang/moostr}) package creates the character strings needed for questions of different types and is therefore equivalent to the routines \emph{nm}, \emph{mc} and \emph{sa} in the \emph{moodlequizR} package.

\begin{sloppypar}
Finally the \emph{moodlequiz} package attempts the same as \emph{moodlequiz} (\href{https://github.com/numbats/moodlequiz}{https://github.com/numbats/moodlequiz}), namely creating an xml file that can be imported into moodle. It does so by providing an output format for Rmarkdown documents. It is however much more limited than \emph{moodlequizR} and unfortunately quite badly documented, at least at the time of this writing. Certainly the shiny app in \emph{moodlequizR} provides a much easier start for beginners.
\end{sloppypar}

\hypertarget{pedagogical-considerations}{%
\section{Pedagogical Considerations}\label{pedagogical-considerations}}

The original reason for creating these routines was as a way to lessen, and eventually eliminate, the chore of grading papers. However, not having to grade papers by hand has also changed the way I handle quizzes and homework. Before I would give about 10 homework assignments per semester, each generally covering 2 weeks of classes. With the ability to have Moodle grade the assignments I have switched to about 40 quizzes. These are designed in such a way that a student who understands the material can do a quiz in less than 10 minutes. The quizzes close about 3-4 days after the corresponding material has been discussed in class. The students have generally 3 attempts, and have the opportunity to ask questions before the quiz closes. Because now there is much less time between the class discussion and the assignment, the students appear to have a better retention of the material. It is my impression that this has helped the students to better learn the material.

\hypertarget{basic-workflow}{%
\section{Basic Workflow}\label{basic-workflow}}

The idea of \emph{moodlequizR} is to create a file outside of Moodle that can then be
imported into Moodle. The steps are

\begin{enumerate}
\def\labelenumi{\arabic{enumi}.}
\item
  Create an \textbf{R} script, say \emph{quizxyz.R}, either with the shiny app by running \emph{shinymoodlequizR} or with RStudio.
\item
  Create a file \emph{quizxyz.xml}. This is done automatically by the shiny app, or by running the command
\end{enumerate}

\begin{verbatim}
moodlequizR::make.xml(quizxyz, 20, folder="mylocalfolder")
\end{verbatim}

\begin{enumerate}
\def\labelenumi{\arabic{enumi}.}
\setcounter{enumi}{2}
\item
  Go to Moodle, the Question Bank, and Import \emph{quizxyz.xml}.
\end{enumerate}

An xml file is an extensible markup language file and is one of the standard file types recognized by Moodle. The command above would create 20 versions of the quiz and store a file called \emph{quizxyz.xml} in the folder mylocalfolder.

\hypertarget{interactive-app-shinymoodlequizr}{%
\section{Interactive App shinymoodlequizR}\label{interactive-app-shinymoodlequizr}}

The best way to get started is by using the interactive shiny app included in the package. Start it with

\begin{verbatim}
moodlequizR::shinymoodlequizR()
\end{verbatim}

\begin{figure}[!htbp]
\centering
\includegraphics[width=4in]{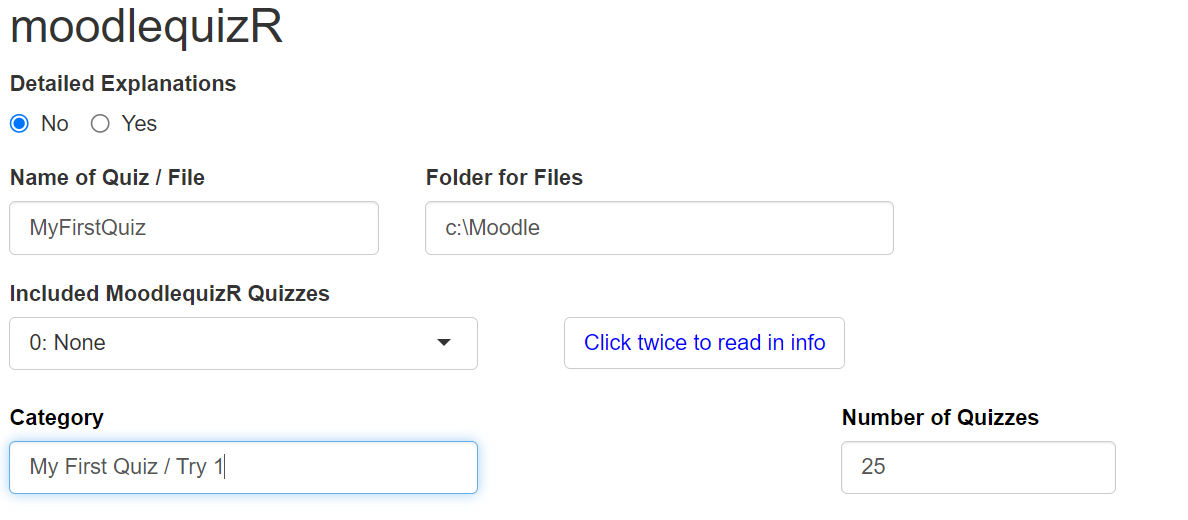}
\caption{Part 1 of a screenshot of the shiny app included in moodlequizR}
\label{fig:fig1}
\end{figure}

Let's say we want to create a quiz where the students are presented with a random data set and are asked to find the mean with \textbf{R}. The data is to be 50 observations drawn from a normal distribution with mean 20 and standard deviation 5, all numbers rounded to 1 digit. After opening the shiny app we can enter the necessary information. See figure~\ref{fig:fig1} to figure~\ref{fig:fig3} for screen shots of the app.

\begin{itemize}
\item Name of Quiz / File: MyFirstQuiz  (this will be the name of the files stored on your computer)  
\item Folder for Files: c:/Moodle (folder where files are to be stored) 
\item Category:  My First Quiz / Try 1 (the category in the Moodle Question Bank where the quiz will appear. For subcategories use /)  
\item Quizzes: 25 (the number of randomized quizzes that should be generated)  
\item  Number of Questions: 1 (the number of questions that the students will have to answer in this quiz.  More generally the number of items that are randomized)
\item Sample size: 50  (this can also be random, 50, 100, 1 means a random sample size from 50 to 100 in steps of 1. For a fixed sample size just use that number)  
\item Type of Data: Normal (a drop down box where we can choose the distribution. Also includes the option R Code, which allows us to generate data from non-standard distributions)  
\item Show data in moodle? Yes (Sometimes we want to hide the raw data)
\item Mean  20  (can be randomize)
\item Std  5  (can be randomize)
\item Round data to 1 digit  (can be randomize)  
\item Sort data? No (choose Yes if data should be sorted)  
\item General Calculations: (any \textbf{R} calculations that might be needed)
\item Start on new line: Yes (should this question appear on the same line as the last one?)  
\item Text of Question: (the text that should appear in Moodle)  
\item R Calculations for this Question: (\textbf{R} commands to find correct answers) 
\item Type of Question: Numeric (here the answer is a number, in other questions it might be multiple choice etc)  
\item Point(s): 100 (how many points is the question worth? This allows for partial credit, so for example we might give 50 points for answers that were not rounded)
\item Precision: 0 (how precise does the answer have to be?)  
\item Text of Answer: (what the student sees once they submitted their solution. If empty it is the same text as the question but often we want to show the student the correct \textbf{R} command)  
\end{itemize}

\begin{figure}[!htbp]
\centering
\includegraphics[width=4in]{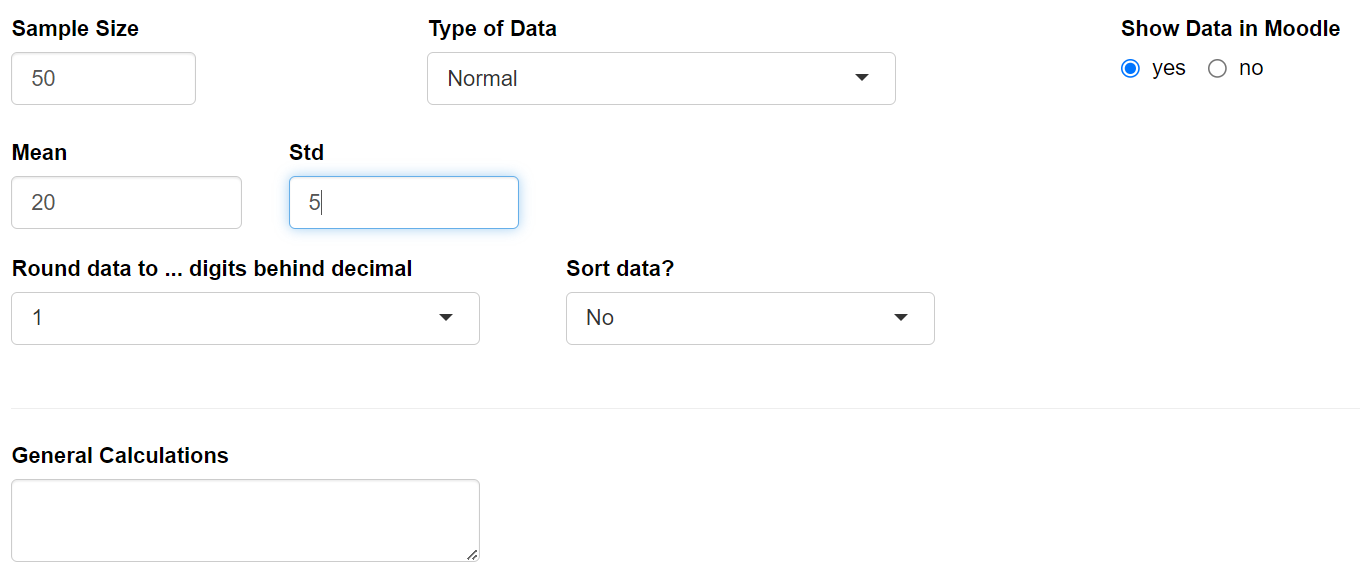}
\caption{Part 2 of a screenshot of the shiny app included in moodlequizR}
\label{fig:fig2}
\end{figure}

\begin{figure}[!htbp]
\centering
\includegraphics[width=4in]{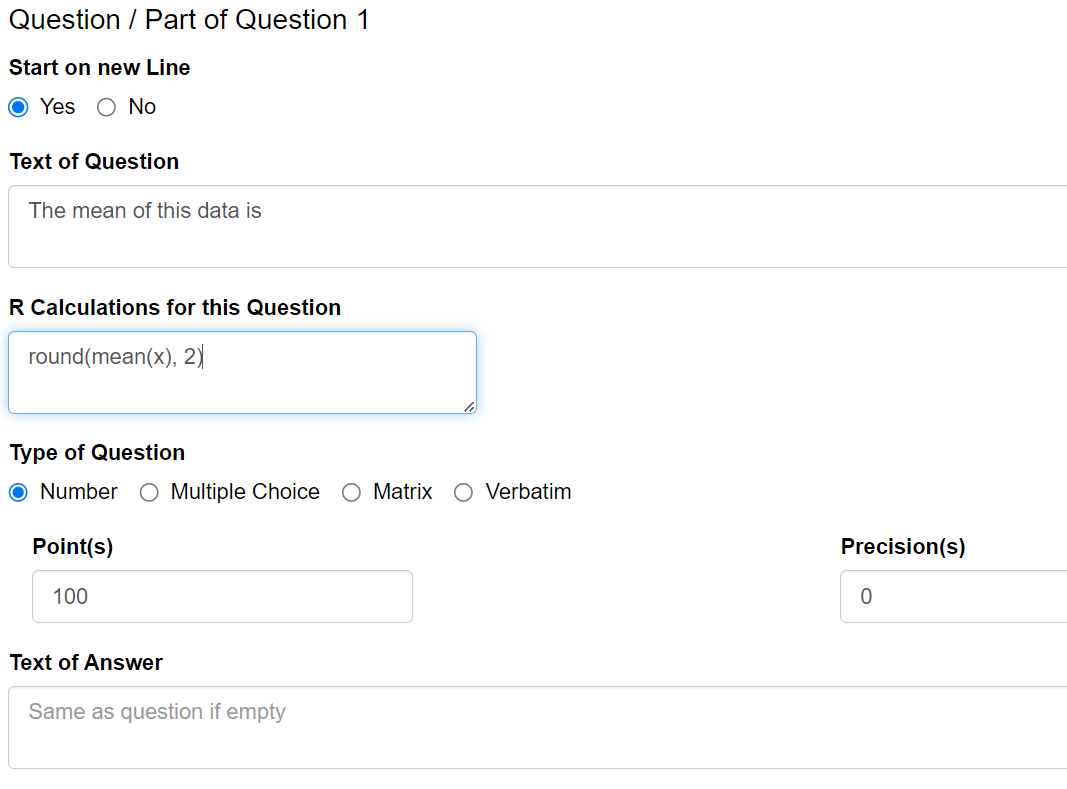}
\caption{Part 3 of a screenshot of the shiny app included in moodlequizR}
\label{fig:fig3}
\end{figure}

Now clicking on Create the Files! at the bottom of the page will create three files:

\begin{enumerate}
\item \emph{MyFirstQuiz.dta}, an \textbf{R} list that can be read into the app the next time it is used, so the information has to be entered only once.
\item \emph{MyFirstQuiz.R}, an \textbf{R} scipt that can be manipulated directly to make further quizzes, see discussion below.
\item \emph{MyFirstQuiz.xml}, a file that can be imported into Moodle.
\end{enumerate}

\begin{figure}[!htbp]
\centering
\includegraphics[width=4in]{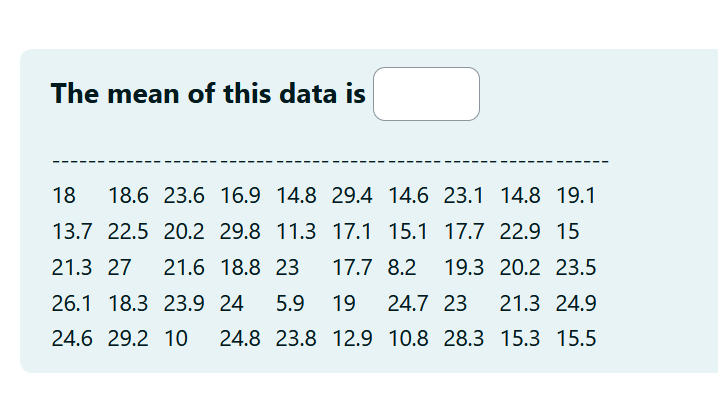}
\caption{Appearance of the quiz in Moodle}
\label{fig:fig4}
\end{figure}

Going to Moodle, importing the XML file into the Question Bank and previewing it shows the screen shot in figure~\ref{fig:fig4}. Now a student can copy the data as usual, switch to \textbf{R} and use the routine \emph{moodlequizR::paste.moodle()} to transfer the data and finally answer the question. 

The routine \emph{moodlequizR::paste.moodle} makes the transfer of data from Moodle to \textbf{R} very easy, and it works on Windows, MacOS and Linux operating systems. In principle transferring data to other programs such as Minitab should also be possible. 

In the app a \emph{Question / Part of Question} is usually just that, a place where the students have to enter their answer to a question. However, a bit more general it is any place where the quiz is randomized, and sometimes that might appear as just text. For an example consider the built-in example 13, where the first Question just displays the output of an R command.

In the text that the user enters in the box the randomized part will appear at the @ symbol, or at the end if no @ is found.

\hypertarget{the-r-script}{%
\section{The \textbf{R} script}\label{the-r-script}}

The heart of \emph{moodlequizR} are \textbf{R} scripts that can be turned into an xml file. These scripts have to have a certain structure. Here is a very simple example of such a script:

\begin{verbatim}
example1 = function() {
  category = "Examples / 1"  # moodle category where quiz will be stored
  quizname = "problem -" # running counter for problems
  n = 50 + sample(0:50, 1)  # randomized sample size
  m = round(runif(1, 90, 110), 1) # randomized mean
  s = round(runif(1, 8, 12), 2) # randomized standard deviation
  d = sample(1:3) # randomized number of digits
  x = rnorm(n, m, s) # generate data
  x = round(x, d) # round it to d digits
  res = round(mean(x), d+1) # calculate correct answer, to one digit more than the data.
  # create question text    
  qtxt =  paste0("<p>The mean of the data is  {1:NM:%100%", res, ":0~%80%", res, ":0.01}</p>" )
  # create answer text with correct answer      
  atxt =  paste0( "<p>The mean of the data is  ", res," </p>" )
  # create hint     
  htxt = "Use the mean command"
  list(qtxt = paste0("<h5>", qtxt, "</h5>", moodle.table(x)),
       htxt = paste0("<h5>", htxt, "</h5>"),
       atxt = paste0("<h5>", atxt, "</h5>"),
       category = category, quizname = quizname)
}
\end{verbatim}

The routine generates a data set of size n (randomly chosen between 50
and 100) from a normal distribution with a
random mean (90-110) and a random standard deviation (8-12) and then rounded to d (1-3) digits. The student
has to find the mean and round to d+1 digits, and gets partial credit (80\%) if the rounding is not done.

The output of the routine has to be a list with five elements:

\begin{enumerate}
\def\labelenumi{\arabic{enumi}.}
\item
  \textbf{category}: this is the category under which moodle will file
  the the problem.
\item
  \textbf{quizname}: the name of the problem (usually just problem - )
\item
  \textbf{qtxt}: the ENTIRE text of the problem as it will appear to the
  students.
\item
  \textbf{htxt}: Hints that students see after their first attempt
\item
  \textbf{atxt}: the ENTIRE text of the answers that the students see
  after the deadline for the quiz has passed.
\end{enumerate}

The \emph{h5} at then end of the routine is html code for \emph{heading 5}. It is used to make the text a bit more readable in Moodle. But more generally we can use any html code and Moodle knows what to do with it. The same is true for latex, as we will see later.

Here is an example with a test that has two questions on the same Moodle page:

\begin{verbatim}
example2 = function() {
   category = "Examples / Percentage 1"
   quizname = "problem -"
# Question 1
   n = sample(200:500, 1)
   p = runif(1, 0.5, 0.6)
   x = rbinom(1, n, p)
   per = round(x/n*100, 1)
   qtxt = paste0("Question 1: In a survey of ", n, " 
      people ", x, " said that they prefer Coca-Cola 
      over Pepsi. So the percentage of people who prefer 
      Coca-Cola over Pepsi is 
      {:NM:%100%", per,":0.1~%80%", per, ":0.5}%")
   htxt = "Don't forget to multiply by 100, don't inlcude % sign in answer"
   atxt = paste0("Question 1: x/n*100 = ", x, "/", n,
          "*100 = ", per, " (rounded to 1 digit)")
# Question 2
   p1 = round(runif(1, 50, 60), 1)
   if(per<p1) mc = c("{:MC:~%100%lower~%0%the same~%0%higher}", " < ")  
   if(p1==per) mc = c("{:MC:~%0%lower~%100%the same~%0%higher}", " = ")
   if(per>p1) mc = c("{:MC:~%0%lower~%0%the same~%100%higher}", " > ")   
   qtxt = paste0(qtxt, "<p>Question 2: In a survey some
     years ago the percentage was ", p1, "%.
    So the percentage now is ", mc[1]) 
   htxt = ""
   atxt = paste0(atxt, "<p>Question 2: ", per , mc[2], p1)

   list(qtxt = paste0("<h5>",qtxt,"</h5>"), 
        htxt = paste0("<h5>", htxt,"</h5>"), 
        atxt = paste0("<h5>", atxt,"</h5>"), 
        category = category, quizname = quizname) 
} 
\end{verbatim}

so one simply creates a long string with all the questions and another
with all the answers.

\hypertarget{embedded-answer-cloze-question-types}{%
\section{Embedded Answer (CLOZE) Question Types}\label{embedded-answer-cloze-question-types}}

Moodle has a number of formats for writing questions. \emph{moodlequizR} uses the CLOZE format because it has a number of features that work well for our purpose. There are three basic question types:

\hypertarget{multiple-choice-questions}{%
\subsection{Multiple choice questions}\label{multiple-choice-questions}}

\textbf{Example}: For this data set the mean is \{1:MC:\(\sim\)\%0\%lower\(\sim\)\%0\%not equal to\(\sim\)\%100\%higher\} than the median.

In this case the students are presented with a drop-down box with the
three options: lower, not equal to and higher. higher is the correct answer, so it gets 100\%, the others get 0\%. One
can also give (say) 65\% for partial credit. The 1 in the front means
the problem is worth 1 point.

\emph{moodlequizR} includes some routines that make writing these questions easier. The routine \emph{mc} creates the code needed for multiple
choice questions. It has a number of standard choices already
implemented. It is run with \emph{moodlequizR::mc(options, w)}. Here options is a character vector with the choices (or a number for some common ones that are predefined), w is the vector of percentages, usually 100 for the correct answer and 0 for the others.
So the following command creates the text for the multiple choice question above:

\begin{verbatim}
moodlequizR::mc(2, c(0,0, 100))
\end{verbatim}

Notice that the output of this routine is a list with both the question and the answer text.

The built in options are

\begin{enumerate}
\def\labelenumi{\arabic{enumi}.}
\item
  lower, not equal to, higher, can't tell
\item
  lower, not equal to, higher
\item
  is statistically significant, is not statistically significant
\item
  is statistically significant, is not statistically significant, can't tell
\item
  is, is not
\item
  Male, Female
\item
  true, false
\item
  has, does not have
\item
  \(\ne, <, >\), can't tell
\item
  \(\ne, <, >\)
\item
  \(\mu, \pi, \sigma, \lambda, \rho\), other
\end{enumerate}

\hypertarget{numerical-answer-questions}{%
\subsection{Numerical answer questions}\label{numerical-answer-questions}}

\textbf{Example}: The mean is \{2:NM:\%100\%54.7:0.1\(\sim\)\%80\%54.7:0.5\}.

Now the students see a box and have to type in a number.

Here 54.7 is the correct answer, which get's 100\%. Any answer within
\(\pm 0.1\) is also correct, allowing for rounding to 1 digit. The
answer 55 gets 80\% (a bit to much rounding). The question is worth 2 points.

The routine \emph{moodlequizR::nm} creates numerical questions. The format is \emph{nm(x, w, eps, ndigits, pts=1)}.

x is a vector of possible answers, w is the vector of percentages and
eps is a vector of acceptable range \(\pm\). Alternatively one can use
the argument ndigits. With (say) ndigits=1 the answer has to be rounded
to one digit behind the decimal, all other roundings get partial credit.

So for the above example we need to run

\begin{verbatim}
nm(54.7, c(100, 80), c(0.1, 0.5))
\end{verbatim}

If just one number is correct, say 50, and the answer has to be given
exactly, use \emph{nm(50)}.

\hypertarget{text-answers}{%
\subsection{Text answers}\label{text-answers}}

\emph{Example}: The correct method for analysis is the \{1:SA:correlation coefficient\}

Here the student sees a box and has to type in some text.

There is also \{1:SAC:correlation coefficient\} if the case has to
match.

This type of problem has the issue of empty spaces. So if the student
typed correlation coefficient with two spaces it would be judged wrong.
The solution is to use *s: \{1:SA:*correlation*coefficient*\}.

To create text questions we can use \emph{moodlequizR::sa}. The format is
\emph{sa(txt, w=100, caps=TRUE, pts=1)}. This function automatically adds * in a number of places. txt can be a
vector. As an example the above text question can be created with

\begin{verbatim}
sa("correlation coefficient")
\end{verbatim}

With these routines example2 can be re-written as follows:

\begin{verbatim}
example2   = function() {
   category = "moodlequizR / Percentage 2"
   quizname = "problem -"
 
# Question 1

n = sample(200:500, 1)
   p = runif(1, 0.5, 0.6)
   x = rbinom(1, n, p)
   per = round(x/n*100, 1)
   qtxt = paste0("Question 1: In a survey of ", n, "
       people ", x, " said that they prefer Coca-Cola over
       Pepsi. So the percentage of people who prefer
        Coca-Cola over Pepsi is ", 
      nm(per, c(100, 80), c(0.1, 0.5)), "%")
   htxt = "Don't forget to multiply by 100, don't inlcude % sign in answer"
   atxt = paste0("Question 1: x/n*100 = ", x, "/", n, "*100 = ", per, " (rounded to 1 digit)")

# Question 2

   p1 = round(runif(1, 0.5, 0.6)*100, 1)
   if(per<p1) {w = c(100, 0, 0); amc ="<"}
   if(per==p1) {w = c(0, 100,  0); amc ="="}
   if(per>p1) {w = c(0, 0, 100); amc =">"}
   opts = c("lower", "the same", "higher")  
   
   qtxt = paste0(qtxt, "<p>Question 2: In a survey some
        years ago the percentage was ", p1, "%.
        So the percentage now is ", mc(opts, w)) 
   htxt = ""
   atxt = paste0(atxt, "<p>Question 2: ", per , amc, p1)

   list(qtxt = paste0("<h5>",qtxt,"</h5>"), 
        htxt = paste0("<h5>", htxt,"</h5>"), 
        atxt = paste0("<h5>", atxt,"</h5>"), 
        category = category, quizname = quizname) 
}
\end{verbatim}

\hypertarget{creating-the-xml-file}{%
\section{Creating the xml file}\label{creating-the-xml-file}}

Once the \textbf{R} script is written it can be read into \textbf{R} and then the \emph{make.xml} routine can be used to create the Moodle input file
\emph{quizxyz.xml}. This is done with the command

\begin{verbatim}
moodlequizR::make.xml(quizxyz, 20, folder="mylocalfolder")
\end{verbatim}

which creates 20 quizzes and stores the file \emph{quizxyz.xml} in the folder \emph{mylocalfolder}. The routine can also pass arguments to the \textbf{R} script.

Next the user has to open Moodle and import this file.

Moodle has two Import places. One is for importing
existing questions from other courses. The user needs to go to Questions - Import, select XML, drop quizxyz.xml into the box and hit enter.

\hypertarget{multiple-stories}{%
\section{Multiple Stories}\label{multiple-stories}}

In statistics we usually use word problems. So how can we do that?
Essentially we can make up a couple of stories:

\begin{verbatim}
example3 = function(whichstory) {
   if(missing(whichstory)) whichstory=sample(1:3, 1)
   category = paste0("moodlequizR / Percentage : Story : ", whichstory)
   quizname = "problem -"
    
   if(whichstory==1) {  
      n = sample(200:500, 1)
      p = runif(1, 0.5, 0.6)
      x = rbinom(1, n, p)
      per = round(x/n*100, 1)
      qtxt = paste0("In a survey of ", n, " people ", x, " said that they prefer Coca-Cola over Pepsi. 
			              So the percentage of people who prefer Coca-Cola over Pepsi is ", 
										nm(per, c(100, 80), c(0.1, 0.5)), "%")
   } 
   if(whichstory==2) {  
      n = sample(1000:1200, 1)
      p = runif(1, 0.45, 0.55)
      x = rbinom(1, n, p)
      per = round(x/n*100, 1)
      qtxt = paste0("In a survey of ", n, " likely voters ", x, " said that they would vote for 
			              candidate A. So the percentage of people who will vote for candidate A is ", 
									  nm(per, c(100, 80), c(0.1, 0.5)), "%")      
   }
   if(whichstory==3) {  
      n = sample(100:200, 1)
      p = runif(1, 0.75, 0.95)
      x = rbinom(1, n, p)
      per = round(x/n*100, 1)
      qtxt = paste0("In a survey of ", n, " people ", x, " said that they are planning a vacation 
			               this summer. So the percentage of people who are planning a vacation is ", 
										 nm(per, c(100, 80), c(0.1, 0.5)), "%")      
     }
   htxt = ""
   atxt = paste0("x/n*100 = ", x, "/", n, "*100 = ", 
          per, " (rounded to 1 digit)")

   list(qtxt = paste0("<h5>",qtxt,"</h5>"), 
        htxt = paste0("<h5>", htxt,"</h5>"), 
        atxt = paste0("<h5>", atxt,"</h5>"), 
        category = category, quizname = quizname) 
}
\end{verbatim}

Note that we can match each story with likely numbers, so in the
Coca-Cola vs Pepsi story the sample size n is between 200 and 500 whereas in the votes story it is between 1000 and 1200.

By default the routine above creates about equally many of the three stories. If we wish to just create those with (say) story 2 we can run

\begin{verbatim}
make.xml(example3, 20, whichstory=2)
\end{verbatim}

\hypertarget{data-sets}{%
\section{Data Sets}\label{data-sets}}

Often in Statistics we have data sets the students need to use. So these data sets need to be displayed properly in the quiz and it must be easy for the students to transfer them to \textbf{R}.

To display the data in the quiz we have the \emph{moodle.table} function. If
it is called with a vector it arranges it as a table with (ncol=) 10
columns. If it is called with a matrix or data frame it makes a table as is.

To get the data from the quiz into \textbf{R} \emph{moodlequizR} has the routine \emph{paste.data}. All the students have to do is highlight the data
(including column names if present) in the quiz with the mouse, right
click copy, switch to \textbf{R} and run

\begin{verbatim}
moodledata = paste.data()
\end{verbatim}

There will now be an object called \emph{moodledata} in \textbf{R}. If it is a single vector it can be used as is, say

\begin{verbatim}
mean(moodledata)
\end{verbatim}

If the data was a table with several columns \emph{moodledata} is a dataframe. The routine correctly reads vectors, both numeric and character. Also tables with a mix of character and numeric columns. It works on both Windows and Apple operating systems.

So we could have the following test:

\begin{verbatim}
example4 = 
function() {
   category = "moodlequizR / Percentage from Raw Data" 
   quizname = "problem -"

   n = sample(200:500, 1)
   p = runif(1, 0.5, 0.6)
   x = sample(c("Coca-Cola", "Pepsi"), size=n, replace=TRUE, prob=c(p,1-p))
   per = round(table(x)[1]/n*100, 1)

   qtxt = paste0("In a survey people were asked whether they prefer Coca-Cola over Pepsi. Their answers are below. So the percentage of people who prefer Coca-Cola over Pepsi is ", nm(per, c(100, 80), c(0.1, 0.5)), "%<hr>",
    moodle.table(x))
   htxt = "" 
   atxt = paste0("x/n*100 = ", table(x)[1], "/", n, "*100 = ", per, " (rounded to 1 digit)")
   
   list(qtxt = paste0("<h5>",qtxt,"</h5>"), 
        htxt = paste0("<h5>", htxt,"</h5>"),
        atxt = paste0("<h5>", atxt,"</h5>"), 
        category = category, quizname = quizname) 
}
\end{verbatim}

Here the students see several lines of text Coca-Cola, Coca-Cola, Pepsi etc. After copy-pasting this to \textbf{R} they can use the table command to find the answer.

\hypertarget{graphs}{%
\section{Graphs}\label{graphs}}

Often in statistics we use graphs. We can use them as part of a Moodle quiz as follows: first
one needs to install the \textbf{R} package \textbf{base64}.

Say we want the quiz to show a histogram and the student has to decide
whether or not it is bell-shaped:

\begin{verbatim}
example5 = function(bell=TRUE) {
   require(base64)
     require(ggplot2)
   category = paste0("moodlequizR / 
      bell shaped ? ", ifelse(bell, "Yes", "No")) 
   quizname = "problem -"

   n = 1000
   if(bell) x = rnorm(n, 10, 3)
   else x= rchisq(n, 2) + 8
   
   plt = ggplot(data = data.frame(x = x), aes(x)) + geom_histogram(aes(y = ..density..), color = "black", fill = "white", binwidth = 0.2)
   plt64 = moodlequizR::png64(plt)
   if(bell) mmc = mc(5, c(100, 0))[[1]]
   else mmc = mc(5, c(0, 100))[[1]]
   qtxt = paste0("The data shown in this histogram ", mmc, " bell shaped<hr>", plt64)   
   if(bell) atxt = "It is bell shaped"
   else atxt = "It is not bell shaped"
   htxt = "" 

   list(qtxt = paste0("<h5>", qtxt, "</h5>"), 
        htxt = paste0("<h5>", htxt, "</h5>"),         
        atxt = paste0("<h5>", atxt, "</h5>"), 
        category = category, quizname = quizname) 
}
\end{verbatim}

\hypertarget{non-statistics-courses}{%
\section{Non Statistics Courses}\label{non-statistics-courses}}

The same basic ideas can be used to make random quizzes and exams for other
courses. Here is an example of a quiz for a calculus class. The students have to find the integral

\[\int_{A}^{B} xe^x dx\]

where A and B are chosen randomly.

\begin{verbatim}
example6 = function() {
   category = "moodlequizR / Integral"
   quizname = "problem -"

   A = round(runif(1, 0, 1), 1)
   B = round(runif(1, 1, 2), 1)
   fun = function(x) {x*exp(x)}
   I = round(integrate(fun, A, B)[[1]], 2)
   qtxt = paste0("\\(\\int_{", A, "}^{", B, "} xe^x dx = \\)", nm(I, eps=0.1))
   htxt = ""
   atxt = paste0("\\(\\int_{", A, "}^{", B, "} xe^x dx = //)", I)
   list(qtxt = paste0("<h5>", qtxt, "</h5>"), 
        htxt = paste0("<h5>", htxt, "</h5>"),         
        atxt = paste0("<h5>", atxt, "</h5>"), 
        category = category, quizname = quizname) 
}
\end{verbatim}

Here we see that one can use latex notation in Moodle quizzes, and so display formulas!

\hypertarget{question-type-formula}{%
\section{Question type formula}\label{question-type-formula}}

Often we would like a test to have a second question in such a way that it depends on the students answer to the first question. As a simple example, say the first question requires the student to find the p value of a hypothesis test, and the second question asks whether or not to reject the null hypothesis at the 5\% level. Let's assume that the correct answers are \(p=0.023\) and Yes. Now say a student makes a mistake and finds a p value of \(0.076\). They should then of course choose No on the second question, but if they do so they will receive 0 points.

Moodle provides a somewhat cumbersome way of creating such follow-up questions with the correct answer depending on their previous answer(s). This is done using the formula question type. While this is not implemented in the shiny app it can be done using \emph{moodlequizR} by editing the \textbf{R} script directly.

\hypertarget{included-example-quizzes}{%
\section{Included Example Quizzes}\label{included-example-quizzes}}

\emph{moodlequizR} has 15 included example quizzes. These can be called up in the shiny app and used both as illustrations on what can be done and as starting points for more complicated quizzes and exams. In the following we provide short descriptions of them:

\begin{enumerate}
\def\labelenumi{\arabic{enumi}.}
\item
  \textbf{Mean} A simple first example where students have to find the mean of a data set. Sample size is chosen randomly from 50 to 100. Data comes from a normal distribution with random mean and standard deviation, rounded to one digit behind the comma. Answer has to be rounded to two digits, otherwise student gets partial credit (80\%).
\item
  \textbf{Mean and Median} Data is the same as in example 1. Students have to find the mean, the median and decide which of the two is larger. This illustrates the use of multiple choice questions.
\item
  \textbf{Five Number Summary} Data is from a Beta distribution with randomly chosen sample size. Data is then rounded to either 1, 2 or 3 digits. Students have to find the five number summary. This example illustrates the use of the General Calculations box to do the needed calculations and shows how to have a table in Moodle for the questions and answers.
\item
  \textbf{One Categorical Variable} Students get data for one categorical variable (number of students in a high school) and have to find the percentages. Example illustrates the use of matrices in Moodle.
\item
  \textbf{Two Categorical Variables} Similar to example 4, but now for two categorical variables. Students have to find the column percentages.
\item
  \textbf{Confidence Interval for Mean} Data is as in example 1, students have to find a confidence interval for the mean, with randomly chosen confidence level. Partial credit is given for wrong rounding and wrong confidence level.
\item
  \textbf{Confidence Interval for Percentage} Student needs to find a confidence interval for a percentage. Data here is the number of successes and number of trials.
\item
  \textbf{Hypothesis Testing for Mean} Data is from a normal distribution with randomly chosen mean and standard deviation. Students have to do a hypothesis test. Also illustrates the use of latex code in Moodle quizzes.
\item
  \textbf{Sample Size for Proportion} Students have to find the sample size needed for a confidence interval for a percentage with given error.
\item
  \textbf{Correlation and Regression} Students have to find the correlation coefficient and the regression equation. Illustrates the use of R Code to generate data and how to add a graph to Moodle.
\item
  \textbf{Data downloaded from the internet} Students have to use a provided link to download data from the New York Power Ball	lottery, read the data into R and answer some questions. 
\item
  \textbf{Multiple Stories} An example of how to create quizzes where different students get different stories (from a small list) with the app. As discussed above we recommend to do this via the R script, however.
\item
  \textbf{R output} Here students see the result of an R command and have to answer a number of questions. This uses the routine \emph{moodlequizR::RtoHTML}, which currently can display out put from the commands \emph{t.test} (one and two sample),\emph{binom.test} and \emph{summary(lm))}.  		
\item
  \textbf{Pre-calculus: Solving a Linear System} An example for using \emph{moodlequizR} to generate a quiz for a pre-culculus course. Students have to solve a linear system of equations with numbers randomly chosen from -3,-2,-1, 1, 2, 3. 0 is not allowed to avoid alignment issues in the quiz. This could also be done but requires much more work.
\item
  \textbf{Calculus: Find Derivative and Integral} Example for use in a calculus course. Students are given a function and have to find derivates and integrals.
\end{enumerate}

\hypertarget{conclusions}{%
\section{Conclusions}\label{conclusions}}

The \textbf{R} package \emph{moodlequizR} provides routines that make it easy to create fully randomized quizzes and exams for the Moodle online teaching system, or indeed any system that uses the XML format. The user has a very fine control over what parts are randomized and how the randomization is done. The included shiny app provides a quick and easy way to create such tests even for users with only a passing knowledge of \textbf{R}. The package also includes routines that allow the students to easily transfer data from Moodle to \textbf{R}. The package is available on \href{https://cran.r-project.org/}{CRAN}.

\bibliographystyle{apalike}
\bibliography{references}

\end{document}